\documentstyle[buckow1,psfig,epsfig,12pt]{article}
\begin {document}
\def\titleline{
\hfill {\small HLRZ1998\_4}
\vspace{2cm}
\newtitleline
Renormalization of strongly coupled 
\newtitleline
 U(1) lattice gauge theories
}
\def\authors{
J. Jers\'ak
}
\def\addresses{
Institute of Theoretical Physics E, \\
RWTH Aachen, D-52056 Aachen, Germany
}
\def\abstracttext{
  Recent numerical studies of the 4D pure compact U(1) lattice gauge
  theory, I have participated in, are reviewed.  We look for a
  possibility to construct an interesting nonperturbatively
  renormalizable continuum theory at the phase transition between the
  confinement and Coulomb phases.  First I describe the numerical
  evidence, obtained from calculation of bulk observables on spherical
  lattices, that the theory has a non-Gaussian fixed point. Further
  the gauge-ball spectrum in the confinement phase is presented and
  its universality confirmed. The unexpected result is that, in
  addition to massive states, the theory contains a very light,
  possibly massless scalar gauge ball. I also summarize results of
  studies of the compact U(1) lattice theory with fermion and scalar
  matter fields and point out that at strong coupling it represents a
  model of dynamical fermion mass generation.
  } 
\makefront 

Talk presented at the {\em 31st International Symposium Ahrenshoop on
  the Theory of Elementary Particles,} Buckow, September 2-6, 1997.

\pagebreak
\section{Introduction}
Quantum field theories used in the standard model and its
supersymmetric extensions are either asymptotically free or so-called
trivial theories.  Both are defined in the vicinity of Gaussian fixed
points and can be studied perturbatively. In various collaborations we
address the question whether in four dimensions (4D) there exist
quantum field theories which are not accessible to or anticipated by
perturbation theory. I describe the results of our systematic
numerical study of critical behavior in several compact U(1) gauge
models on the lattice, both pure and with matter fields. Their common
feature is confinement when the bare gauge coupling is strong. It has
never been clear whether this is only a lattice artefact or whether a
confining continuum U(1) gauge theory with confinement can be
constructed. Though not yet conclusive, our results are promising and
we hope to stimulate more theoretical attention.

\section{Renormalization of lattice field theories}
In numerical simulations of lattice field theories the concept of
renormalizability requires the existence of critical behaviour somewhere in
the bare coupling parameter space. Dimensionful observables, e.g. masses $m_s,
m_1, m_2, \dots ,$ $n$-point functions, etc, can be calculated in the lattice
constant units $a$ as functions of couplings in the vicinity of a critical
point or manifold. The dimensionless correlation lengths $\xi_i = 1/am_i
\rightarrow \infty$ when a critical point is approached.

Renormalizability further requires the existence of ``lines of constant
physics'' in the bare coupling parameter space, along which the observables
scale, i.e. their dimensionless ratios $r_1 = m_1/m_s, r_2 = m_2/m_s, \dots ,$
stay (approximately, in practice) constant. If such a line hits the critical
point, one can construct the continuum limit and obtain the values of
dimensionful observables by fixing one mass scale, $m_s$, in physical units.
Then $a = 1/\xi_s m_s \rightarrow 0 $ and $m_i = r_i m_s$. These results are
usually universal, i.e. independent of the detailed choice of the coupling
space and lattice structure, being governed by a limited number of fixed
points.

The lines of constant physics can approach a critical manifold but, before
hitting it, leave the parameter space. Or they can enter phase transitions of
weak first order. In both cases the correlation lengths grow, but stay finite.
Some inherent cut-off $\Lambda = \xi_s^{max} m_s$ remains. If it is large with
respect to the physical scale $m_s$, i.e. if $\xi_s^{max} >> 1$, the theory is
renormalizable in a restricted sense. In this way e.g. the familiar trivial
theories arise.

\section{Pure gauge theory on spherical lattices}


Recently we have reconsidered the oldest candidate for a non-Gaussian fixed
point in the 4D lattice field theory, the phase transition between the
confinement and the Coulomb phases in the pure compact U(1) gauge
theory. We have used the extended Wilson action in order to enlarge the
possibility of finding a second order phase transition,
\begin{equation}\label{ACTION}
         S = -\sum_P w_P
              \left [\beta \cos(\Theta_P) + \gamma
                \cos(2\Theta_P)\right ].
\end{equation}
Here  $w_P = 1$ and $\Theta_P \in [0,2\pi)$ is the plaquette angle,
i.e. the argument of the product of U(1) link variables along a
plaquette $P$. Taking $\Theta_P = a^2gF_{\mu\nu}$, where $a$ is the
lattice spacing, and $\beta + 4\gamma = 1/g^2$, one obtains for weak
coupling $g$ the usual continuum action $S =\frac{1}{4} \int
d^4xF_{\mu\nu}^2$.

Earlier investigations performed as usual on toroidal lattices suggested that
the phase transition, which is clearly of first order at positive $\gamma$,
might be of second order at negative $\gamma$. Though a weak two-state signal
is present there \cite{EvJe85}, it might be a finite size effect. It has been
found that it disappears on lattices with sphere-like topology
\cite{LaNe94b,JeLa96a,JeLa96b,LaPe96}. As this type of lattice does not change
the universality class \cite{HoJa96,HoLa96}, one can use it as well as the
usual toroidal one.

It has turned out that on spherical lattices the transition has
properties typical for a second-order transition. This concerns, in
particular, the dependence on the size of the lattice. Use of modern
finite size scaling (FSS) analysis techniques, and large computer
resources, allowed to determine on finite lattices the critical
properties of the phase transition. A more detailed account of our
work, as well as relevant references, can be found in
Refs.~\cite{JeLa96a,JeLa96b}. Here I list only the most important
findings.

The measurements have been performed at $\gamma = 0, -0.2, -0.5$ on lattices
of the volumina up to about $20^4$.  The FSS behavior of the Fisher zero,
specific heat, some cumulants and pseudocritical temperatures gave consistent
results. The value of the correlation length critical exponent $\nu$ has been
found in the range $\nu = 0.35 - 0.40$.

\begin{figure}[tbp]
  \begin{center}
    \leavevmode
     \psfig{file=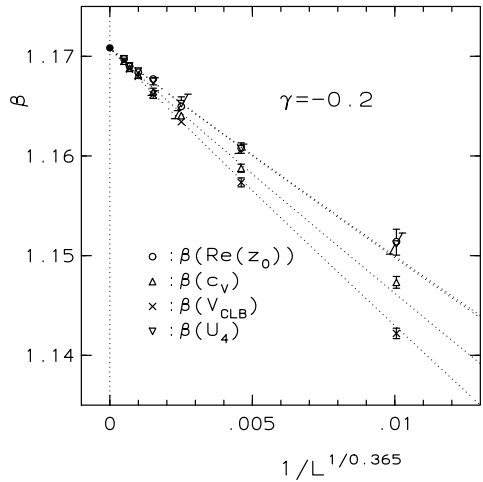,width=7.8cm,angle=180}
       \vspace{-1.7cm}
   \caption{Consistency of the FSS of various pseudocritical
     temperatures with the value $\nu = 0.365$ of the correlation
     length exponent $\nu$ at $\gamma = -0.2$. The figure is from
     Ref.~\protect{\cite{JeLa96b}}}
    \label{fig:3}
  \end{center}
\end{figure}

The most reliable measurement of $\nu$ has been provided by the FSS
analysis of the Fisher zero, i.e. of the first zero $z_0$ of the
partition function in the complex plane of the coupling $\beta$. The
expected behavior with increasing volume $V$ is $ \mbox{Im}\,z_0
\propto V^{-1/D\nu}$.  The joint fit to the data at all three $\gamma$
values gives $ \nu = 0.365(8)$. The scaling behavior of various
pseudocritical temperatures, which have been determined from several
other observables, is consistent with this value, as seen in
Fig.~\ref{fig:3}.

The consistency of the data with the FSS theory suggests that the
phase transition is of second order, as it implies growing correlation
lengths.  Unfortunately, it does not exclude that this growth stops on
lattices substantially larger than those we have used. Very weak first
order phase transition is still possible \cite{DeHa88,Ha88}, which
would mean that the cut-off cannot be made really infinite. The
scaling behaviour could be governed by a fixed point ``behind'' the
phase transition.

In a very recent paper (which appeared after the Symposium)
\cite{CaCr98} Campos et al. suggest that this is what happens. Of
course, also their data require an extrapolation to the infinite
volume. For example latent heat, associated with the two-state signal
present on toroidal lattices, decreases with volume. It can be
extrapolated both to nonvanishing values and zero, depending on the
ansatz. At some second-order transitions, two-state signals due to
finite size effects are known to vanish extremely slowly with volume,
mocking up a first order. It is most probably not possible to clarify
the issue beyond any doubt by numerical methods. But the theory might
be interesting even if the cut-off could not be made infinite but only
much larger than the physical scale, as in physical applications we
are dealing with effective theories anyhow.
%

\section{Gauge-ball spectrum in the confinement phase}

In any case the scaling behaviour of as many observables as possible should be
determined.  Therefore we have investigated at $\gamma = -0.2$ and $-0.5$ the
spectrum of the theory in the confinement phase \cite{CoFr97a,CoFr98a}. It
consists of massive ``gauge balls'', states analogous to glue balls in pure
QCD. In particular, no massless photon is present in this phase. Also the
string tension $\sigma$ has been estimated.

\begin{figure}[tbp]
  \begin{center}
    \psfig{file=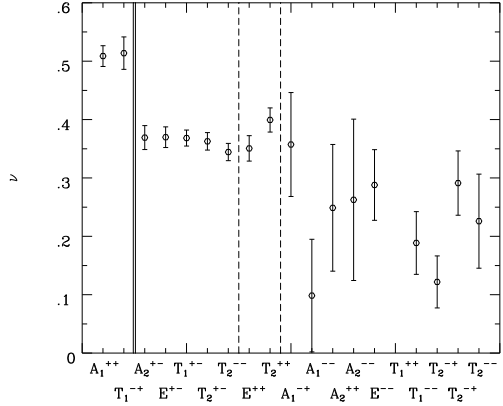,angle=90,width=7.8cm,bbllx=50,
      bblly=110,bburx=545,bbury=774}
  \end{center}
  \caption{Values of $\nu_j$ obtained at $\gamma = -0.2$ in each
    gauge-ball channel separately. The double vertical line separates
    two groups with distinctly different $\nu_j$. The results to the
    right of the dashed vertical lines correspond to heavier states
    which are very difficult to measure. The figure is from
    Ref.~\protect{\cite{CoFr97b}}}
  \label{fig:nu}
\end{figure}

The gauge-ball masses $m_j$ in various channels $j$ of the cubic group have
been measured at both $\gamma$ for different $\beta$.  Then their scaling
behaviour in the form
\begin{equation}
                    m_j = c_j (\beta_c^j - \beta)^{\nu_j}
\label{FIT_INDIVIDUAL}
\end{equation}
(and similar for $\sqrt\sigma$) has been determined in each channel
$j$ individually. We have found two groups of masses with strikingly
different scaling behaviour. Within the whole $\beta$ range a large
group of the gauge-ball masses scale with roughly the same exponents
$\nu_j$ close to the non-Gaussian value 0.365(8) found in
\cite{JeLa96a,JeLa96b,LaPe96}. Though its accuracy is not yet
satisfactory, the exponent of $\sqrt\sigma$ seems to be consistent
with this value, too. However, in several channels getting
contribution from the $0^{++}$ gauge ball the values of $\nu_j$ are
approximately Gaussian, i.e.  $1/2$. This is shown in
Fig.~\ref{fig:nu}.  The values $\beta_c^j$ are quite consistent with
each other in all channels for each $\gamma$.  Therefore, in the
further analysis we have assumed the same value of $\beta_c$ in all
channels.
%
%

Assuming the same exponent $\nu_j$ for each group, denoted $\nu_{\rm ng}$ and
$\nu_{\rm g}$ for the non-Gaussian and Gaussian group, respectively, a joint
fit was performed with these two exponents, common $\beta_c$, and the
individual amplitudes $c_j$ as free parameters:
\begin{equation}
                    m_j = c_j \tau^{\nu_{\rm f}}, \quad {\rm f = ng, g}
\label{FIT_GLOBAL}
\end{equation}
The resulting values of the exponents at $\gamma = -0.2$ are
\begin{eqnarray}
                \nu_{\rm ng} &=&  0.367(14)             \nonumber \\
                \nu_{\rm g}  &=&  0.51(3).
\label{NUGLOB}
\end{eqnarray}
The preliminary results at $\gamma = -0.5$ are consistent with these
values \cite{CoFr98a}.
%
\begin{figure}[tbp]
  \begin{center}
    \leavevmode
    \psfig{file=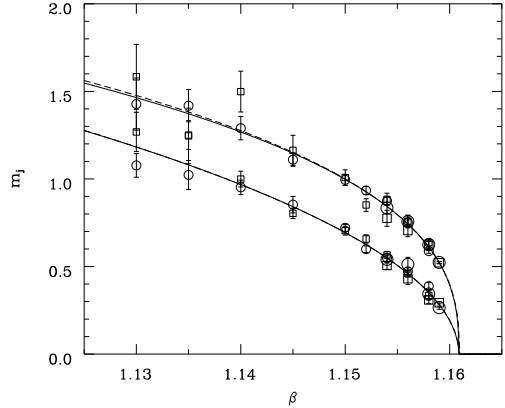,angle=90,width=7.8cm,bbllx=50,
      bblly=110,bburx=545,bbury=774}
  \end{center}
  \caption{Masses of the $0^{++}$ (lower points) and $1^{+-}$
    glue balls in the confinement phase versus $\beta$ at $\gamma =
    -0.2$. The curves are fits \protect{(\ref{FIT_GLOBAL})} with the
    $\nu$ values \protect{(\ref{NUGLOB})}.  The figure is from
    Ref.~\protect{\cite{CoFr97b}}.}
  \label{fig:2scale}
\end{figure}
%

From these results we conclude that the system has two mass scales which we
denote by $m_{\rm ng}$ and $m_{\rm g}$. In Fig.~\ref{fig:2scale} we
show their behaviour. Except $0^{++}$, the gauge-ball masses and presumably
$\sqrt\sigma$ scale proportional to $m_{\rm ng}$. The ratios $r_j = m_j/m_{\rm
  ng}$ are thus constant. This holds for both $\gamma$ values. Thus, within
the limits of numerical determination, we have found two lines of constant
physics. We have further found that $r_j$ are independent of $\gamma$, which
indicates universality. This suggests that the scaling behaviour of the pure
U(1) gauge theory belongs at different $\gamma$ to the same universality class
governed by one non-Gaussian fixed point. The parameter $\gamma$ is irrelevant
in this class as long as it is kept negative.

Choosing $m_{\rm ng}$ for a physical scale, we can consider a continuum theory
(or a theory with large cut-off) with the spectrum given by the values of
$r_j$. They are given in \cite{CoFr97a}. As $m_{\rm g}/m_{\rm ng}$ approaches
zero, this theory would contain massless (or very light) scalar, which could
possibly decouple.

What kind of theory would it be? The procedure we have applied to the
spectrum and $\sigma$ can be extended to all observables and, in
general, to $n$-point functions. If scaling in corresponding powers of
$m_{\rm ng}$ is found, one can, in principle, construct the theory
without having any continuum Lagrangian for it. Or it could be some
non-polynomial Lagrangian in the fields corresponding to the states in
the spectrum. As $\sigma$ would possibly be finite and nonvanishing in
the $m_{\rm ng}^2$ units, it would be a confining theory.  An
interesting proposal for an approximate continuum Lagrangian has been
made in Ref.~\cite{AmEs97} and reported at this workshop by
N.~Sasakura.

The pure U(1) lattice gauge theory with the Villain (periodic Gaussian) action
belongs to the same universality class \cite{LaPe96}.  Rigorous dual
relationships valid for that action imply that also the following 4D models
are governed by the same non-Gaussian fixed point: the Coulomb gas of monopole
loops \cite{BaMy77}, the noncompact U(1) Higgs model at large negative squared
bare mass (frozen 4D superconductor) \cite{Pe78,FrMa86}, and an effective
string theory equivalent to this Higgs model \cite{PoSt91,PoWi93}.

\section{Compact U(1) gauge theory with matter fields}

These findings raise once again the question, whether in strongly
interacting 4D gauge field theories further non-Gaussian fixed points
exist, which might possibly be of interest for theories beyond the standard
model. The pursuit of this question requires an introduction of matter
fields. Therefore we have investigated some extensions of the pure compact
U(1) gauge theory which might have interesting fixed points. We introduce
fermion and scalar matter fields of unit charge, either each separately
or both simultaneously. Because of space limits I give only a very brief
overview of the results.

The compact fermionic QED has some analogous promising properties in the
quenched approximation \cite{CoFr98b}. In compact scalar QED we confirm the
Gaussian behavior at the endpoint of the Higgs phase transition line
\cite{FrJe98b}. In a theory with both scalar and fermion matter fields ($\chi
U \phi$ model), the chiral symmetry is broken, and the mass of unconfined
composite fermions $F = \phi^{\dagger}\chi$ is generated dynamically
\cite{FrJe95a}. It might be an explanation or alternative to the Higgs-Yukawa
sector of the SM.  In 2D \cite{FrJe96c} and 3D \cite{BaFr98} it belongs very
probably to the universality class of the 2D and 3D Gross-Neveu model,
respectively, and is thus nonperturbatively renormalizable. In the 4D $\chi U
\phi$ model we demonstrate the existence of a tricritical point where the
scaling behaviour is distinctly difrerent from the four-fermion theory,
enhancing the chances for renormalizability \cite{FrJe95a,FrJe98a,FrJe98b}.
Here, apart from the massive fermion and Goldstone boson, the spectrum
contains also a massive $0^{++}$ gauge ball. The particular role of this state
both in pure U(1) and in the $\chi U \phi$ model is remarkable and needs a
theoretical explanation.

I thank J. Cox, W. Franzki, C. B. Lang, T. Neuhaus, A. Seyfried, and P.W.
Stephenson for collaboration on the topics I have described, and J.
Ambjorn, I. Campos, and H.  Pfeiffer for discussions. The computations
have been performed on the computers of HLRZ J\"ulich, RWTH Aachen,
and KFU Graz. Our work was supported by Deutsches BMBF and DFG.


\bibliographystyle{wunsnot}   

\begin{thebibliography}{10}

\bibitem{EvJe85}
H.~G. Evertz, J.~Jers\'ak, T.~Neuhaus{,}{ and }\relax P.~M.
  Zerwas, {\em Nucl. Phys.\/} {\bf B251 [FS13]} (1985) 279.

\bibitem{LaNe94b}
C.~B. Lang{ and }\relax T.~Neuhaus, {\em Nucl. Phys.\/}
  {\bf B431} (1994) 119.

\bibitem{JeLa96a}
J.~Jers\'ak, C.~B. Lang{,}{ and }\relax T.~Neuhaus,
  {\em Phys. Rev. Lett.\/} {\bf 77} (1996) 1933.

\bibitem{JeLa96b}
J.~Jers\'ak, C.~B. Lang{,}{ and }\relax T.~Neuhaus,
  {\em Phys. Rev.\/} {\bf D54} (1996) 6909.

\bibitem{LaPe96}
C.~B. Lang{ and }\relax P.~Petreczky, {\em Phys. Lett.\/}
  {\bf 387B} (1996) 558.

\bibitem{HoJa96}
C.~Hoebling, A.~Jakovac, J.~Jers\'ak, C.~B. Lang{,}{ and }\relax
  T.~Neuhaus, {\em Nucl. Phys. B (Proc. Suppl.)\/} {\bf 47}
  (1996) 815.

\bibitem{HoLa96}
C.~Hoebling{ and }\relax C.~B. Lang, {\em Phys. Rev.\/}
  {\bf B54} (1996) 3434.

\bibitem{DeHa88}
K.~Decker, A.~Hasenfratz{,}{ and }\relax P.~Hasenfratz,
  {\em Nucl. Phys.\/} {\bf B295 [FS21]} (1988) 21.

\bibitem{Ha88}
A.~Hasenfratz, {\em Phys. Lett.\/} {\bf 201B} (1988) 492.

\bibitem{CaCr98}
I.~Campos, A.~Cruz{,}{ and }\relax A.~Taranc{\'o}n,
  {\em First order signatures in 4D pure compact U(1) gauge theory with toroidal and spherical topologies\/}, hep-lat/9711045.

\bibitem{CoFr97a}
J.~Cox, W.~Franzki, J.~Jers\'ak, C.~B. Lang,
  T.~Neuhaus{,}{ and }\relax P.~W. Stephenson, {\em Nucl. Phys. B (Proc. Suppl.)\/} {\bf 53} (1997) 696.

\bibitem{CoFr98a}
J.~Cox{ et~al.}, {\em Scaling of gauge balls and static potential in the confinement phase of the pure U(1) lattice gauge theory\/}, HLRZ1997\_40,
  hep-lat/9709054.

\bibitem{CoFr97b}
J.~Cox, W.~Franzki, J.~Jers\'ak, C.~B. Lang,
  T.~Neuhaus{,}{ and }\relax P.~W. Stephenson, {\em Nucl. Phys.\/} {\bf B499} (1997) 371.

\bibitem{AmEs97}
J.~Ambj{\o}rn, D.~Espriu{,}{ and }\relax N.~Sasakura,
  {\em U(1) lattice gauge theory and $N=2$ supersymmetric Yang-Mills theory\/}, hep-th/9707095.

\bibitem{BaMy77}
T.~Banks, R.~Myerson{,}{ and }\relax J.~Kogut, {\em Nucl. Phys.\/} {\bf B129} (1977) 493.

\bibitem{Pe78}
M.~E. Peskin, {\em Ann. Phys.\/} {\bf 113} (1978) 122.

\bibitem{FrMa86}
J.~Fr{\"o}hlich{ and }\relax P.~A. Marchetti, {\em Europhys. Lett.\/}
  {\bf 2} (1986) 933.

\bibitem{PoSt91}
J.~Polchinski{ and }\relax A.~Strominger, {\em Phys. Rev. Lett.\/}
  {\bf 67} (1991) 1681.

\bibitem{PoWi93}
M.~I. Polikarpov, U.-J. Wiese{,}{ and }\relax M.~A. Zubkov,
  {\em Phys. Lett.\/} {\bf 309B} (1993) 133.

\bibitem{CoFr98b}
J.~Cox, W.~Franzki, J.~Jers\'ak, C.~B. Lang{,}{ and }\relax
  T.~Neuhaus, {\em Strongly coupled compact lattice QED with staggered fermions\/}, PITHA 97/22, HLRZ 19/97, hep-lat/9705043.

\bibitem{FrJe98b}
W.~Franzki{ and }\relax J.~Jers\'ak, {\em Dynamical fermion mass generation at a tricritical point in strongly coupled U(1) lattice gauge theory\/}, PITHA 97/43, HLRZ1997\_66, hep-lat/9711039.

\bibitem{FrJe95a}
C.~Frick{ and }\relax J.~Jers\'ak, {\em Phys. Rev.\/} {\bf D52}
  (1995) 340.

\bibitem{FrJe96c}
W.~Franzki, J.~Jers\'ak{,}{ and }\relax R.~Welters,
  {\em Phys. Rev.\/} {\bf D54} (1996) 7741.
  
\bibitem{BaFr98} 
  I.~M. Barbour, W.~Franzki{,}{ and }\relax N.~Psycharis,
  contribution to the conference Lattice '97; \\
  I.~M. Barbour, E.~Focht, W.~Franzki{,} J.~Jers\'ak, { and }\relax
  N.~Psycharis, {\em Strongly coupled U(1) lattice gauge theory with
    dynamical fermion mass generation in three dimensions}, in
  preparation.

\bibitem{FrJe98a}
W.~Franzki{ and }\relax J.~Jers\'ak, {\em Strongly coupled U(1)
  lattice gauge 
 theory as a microscopic model of Yukawa theory\/}, PITHA 97/42,
  HLRZ1997\_65, hep-lat/9711038.

\end{thebibliography}


\end{document}